\documentclass[letterpaper,conference]{IEEEtran}
\usepackage{epsfig}
\usepackage{times}
\usepackage{float}
\usepackage{afterpage}
\usepackage{amsmath}
\usepackage{amstext}
\usepackage{amssymb,bm}
\usepackage{latexsym}
\usepackage{color}
\usepackage{graphicx}
\usepackage{amsmath}
\usepackage{amsthm}
\usepackage{graphicx}
\usepackage[center]{caption}
\usepackage{pstricks}
\usepackage{subfigure}
\usepackage{booktabs}
\usepackage{enumerate}

\newtheorem{theorem}{Theorem}

\newcommand{\Rp}{{R_{1}}}
\newcommand{\Rc}{{R_{2}}}
\newcommand{\Wp}{{W_{1}}}
\newcommand{\Wc}{{W_{2}}}
\newcommand{\Tp}{{T_{1}}}
\newcommand{\Tc}{{T_{2}}}
\newcommand{\Xp}{{X_{1}}}
\newcommand{\Xc}{{X_{2}}}
\newcommand{\Yp}{{Y_{1}}}
\newcommand{\Yc}{{Y_{2}}}
\newcommand{\Zp}{{Z_{1}}}
\newcommand{\Zc}{{Z_{2}}}
\newcommand{\Yf}{{Y_{\mathsf{f}}}}
\newcommand{\Zf}{{Z_{\mathsf{f}}}}

\newcommand{\snr}{\mathsf{S}}
\newcommand{\inr}{\mathsf{I}}
\newcommand{\Sp}{{{\snr}_{1}}}
\newcommand{\Sc}{{{\snr}_{2}}}
\newcommand{\Ip}{{{\inr}_{1}}}
\newcommand{\Ic}{{{\inr}_{2}}}
\newcommand{\Cc}{{\mathsf{C}}}

\newcommand{\gdof}{\mathsf{d}}
\begin{document}

\title{New Outer Bounds for the Interference Channel with Unilateral Source Cooperation} 

\author{%
\IEEEauthorblockN{Martina Cardone$^{\dagger}$, Daniela Tuninetti$^*$, Raymond Knopp$^{\dagger}$, Umer Salim$^{\ddagger}$,}
$^{\dagger}$Eurecom,
Biot, 06410, France, 
Email: \{cardone, knopp\}@eurecom.fr\\
$^*$ University of Illinois at Chicago,
Chicago, IL 60607, USA, 
Email: danielat@uic.edu\\
$^{\ddagger}$ Intel Mobile Communications,
Sophia Antipolis, 06560, France, 
Email: umer.salim@intel.com}\maketitle
\begin{abstract}
This paper studies the two-user interference channel with unilateral source cooperation, which consists of two source-destination pairs that share the same channel and where one full-duplex source can overhear the other source through a noisy in-band link. 
Novel outer bounds of the type $2\Rp+\Rc/\Rp+2\Rc$ are developed for the class of injective semi-deterministic channels with independent noises at the different source-destination pairs. 
The bounds are then specialized to the Gaussian noise case. Interesting insights are provided about when these types of bounds are active, or in other words, when unilateral cooperation is too weak and leaves ``holes'' in the system resources.
\end{abstract}

\section{Introduction}
\label{sec:intro}
A major limitation of current wireless networks is interference. In today's systems, interference is either avoided or treated as noise. Interference avoidance is accomplished by splitting the available time / frequency / space / code resources among the users in such a way that their transmissions are ``orthogonalized''. In practice, perfect user orthogonalization is not possible leading to a residual interference, which is usually treated as noise. This approach may severely limit the system capacity. 
Cooperation among wireless nodes has emerged as a potential technique to enhance performance.
Cooperation leverages the broadcast nature of the wireless medium, i.e., the same transmission can be heard by multiple nodes, thus opening up the possibility that nodes help one another by relaying their message.

Motivated by the potential impact of cooperation in future wireless networks, this paper studies a system consisting of two source-destination pairs that share the same channel. One source, $\mathsf{Tx2}$, overhears the other, $\mathsf{Tx1}$, through a noisy in-band link. Therefore, $\mathsf{Tx2}$, which is here assumed to operate in full-duplex, besides communicating with $\mathsf{Rx2}$, may also allocate some of its resources to boost the crate of $\mathsf{Tx1}$. 
This channel model is referred to as the {\em interference channel with unilateral source cooperation}.
This would model a scenario where, for example, a base station can overhear another base station and consequently help serving its associated mobile users.

\subsection{Related Work} 
Lately, cooperation 
has received significant attention as summarized in what follows.

The Interference Channel (IC) with unilateral source cooperation is a special case of the IC with generalized feedback, or bilateral source cooperation. 
For this network, several outer bounds on the capacity have been derived~\cite{HostMadsenIT06,PVIT11,TuninettiITA10}. A number of schemes, seeking to match these outer bounds, have been developed as well. 
For example, \cite{YANG-TUNINETTI} proposed a strategy 
that exploits rate splitting, superposition coding, partial-decode-and-forward relaying, and Gelfand-Pinsker binning.
This strategy, specialized to the Gaussian noise channel, turned out to match the sum-rate outer bounds of~\cite{TuninettiITA10,PVIT11} to within 19~bits under the assumption of equally strong cooperation links with arbitrary direct and interfering links~\cite{PVIT11} and to within 4~bits in the `strong cooperation regime' with symmetric direct links and symmetric interfering links~\cite{YangHighCoop}.

The IC with unilateral source cooperation was studied in \cite{MirmohseniIT2012}, where it was assumed 
that, at any given time instant, the cooperating source has an a-priori (before transmission begins) access to $L\geq 0$ future channel outputs. 
For the case $L=0$ studied in this paper, \cite{MirmohseniIT2012} derived potentially tighter outer bounds than those in \cite{PVIT11,TuninettiITA10} specialized to unilateral source cooperation. However, these bounds involve several auxiliary random variables 
and it is thus not clear how to evaluate them for the Gaussian noise channel. The authors of \cite{MirmohseniIT2012} also proposed an achievable scheme, whose rate region, as pointed out in \cite[Rem. 2, point 6]{MirmohseniIT2012}, is no smaller than that in \cite[Sec. V]{YANG-TUNINETTI} specialized to unilateral source cooperation. However, in \cite{MirmohseniIT2012}, no performance guarantees in terms of constant gap were given. In \cite{ourICC2014paper}, the capacity of the IC with unilateral source cooperation was characterized to within 2~bits (per user) for a large set of channel parameters that, roughly speaking, excludes the case of weak interference at both receivers. In \cite{ourICC2014paper}, it was pointed out that in weak interference, outer bounds of the type $2\Rp+\Rc/\Rp+2\Rc$ might be necessary to (approximately) characterize the capacity. To the best of our knowledge, this kind of bounds are not available in the literature and their derivation is the main goal of this work.

Source cooperation includes classical feedback as special case. 
\cite{suhtse:ICwithfeedback} determined the capacity to within~2 bits of the IC where each source has perfect output feedback from the intended destination; it showed that $2\Rp+\Rc/\Rp+2\Rc$-type bounds are not needed because output feedback eliminates ``resource holes,'' or system underutilization due to distributed processing captured by the $2\Rp+\Rc/\Rp+2\Rc$ bounds.
In \cite{SahaiIT2013}, the authors studied the symmetric Gaussian channel with all possible output feedback configurations. It showed that the bounds developed in \cite{suhtse:ICwithfeedback} suffice for constant gap characterization except in the case of `single direct feedback link / model (1000).' 
\cite[Theorem IV.1]{SahaiIT2013} proposed a novel outer bound on $2\Rp+\Rc$ for the injective semi-deterministic channel, to capture the fact that one to the second sources does not receive help.
\cite{TandonPoor} characterized the capacity of the `symmetric linear deterministic IC with degraded output feedback'
by developing bounds on $2\Rp+\Rc/\Rp+2\Rc$, whose extension to the Gaussian noise case was left open.
In this work we extend the results of \cite{TandonPoor,SahaiIT2013} to all {\em injective semi-deterministic channels for which, roughly speaking, the noises at the different source-destination pairs are independent.}

\subsection{Contributions and paper organization}
The rest of the paper is organized as follows. 
Section \ref{sec:chmodel} describes the channel model and summarizes known outer bounds. 
Section \ref{sec:outbounds} presents the derivation of novel outer bounds for the Injective Semi-Deterministic IC (ISD-IC) with unilateral source cooperation.
Section \ref{sec:Gaussian} specializes the new bounds to the Gaussian noise channel and,
by using the generalized Degrees-of-Freedom (gDoF) metric, sheds light on when unilateral cooperation enables sufficient coordination among the sources such that the $2\Rp+\Rc/\Rp+2\Rc$-type bounds are not needed to characterize the capacity, i.e., when there are no longer ``resource holes.''

\section{System model and known outer bounds}
\label{sec:chmodel}

A {\em general memoryless IC with unilateral generalized feedback}, or source cooperation, consists of two input alphabets $\left (\mathcal{X}_1,\mathcal{X}_2 \right )$, three output alphabets $\left (\mathcal{Y}_{\mathsf{F2}},\mathcal{Y}_1,\mathcal{Y}_2 \right )$ and a memoryless transition probability $P_{{Y}_{\mathsf{F2}},\Yp,\Yc|\Xp,\Xc}$.
Each $\mathsf{Tx}u$, $u \in \{1,2\}$, has a message $W_{u}\in [1:2^{N R_{u}}]$ for $\mathsf{Rx}u$, where $N$ is the codeword length and $R_u$ is the transmission rate for user $u$ in bits per channel use.
The messages are independent and uniformly distributed on their respective domains. At time $i \in  [1:N]$, $\mathsf{Tx1}$ encodes the message $\Wp$ into $\Xp_i(\Wp)$ and $\mathsf{Tx2}$ maps its message $\Wc$ and its past channel observations into $\Xc_{i}(\Wc,Y_{\mathsf{F2}}^{i-1})$.
At time $N$, $\mathsf{Rx1}$ estimates its intended message $\Wp$ based on all its channel observations as $\widehat{\Wp}(\Yp^N)$, and similarly $\mathsf{Rx2}$ outputs $\widehat{\Wc}(\Yc^N)$.
A rate pair $\left(\Rp,\Rc \right)$ is said to be $\epsilon$-achievable if there exists a sequence of codes indexed by the block length $N$ such that $\max_{u\in\{1,2\}}\mathbb{P}[\widehat{W}_u \neq W_u]  \leq \epsilon$ for some $\epsilon\in[0,1]$. The capacity is the largest non-negative rate region that is $\epsilon$-achievable for any $\epsilon>0$.

The {\em Injective Semi-Deterministic model} (ISD), introduced in \cite{TelatarTseISIT2007} for the IC without cooperation, assumes that the input $\Xp$, resp. $\Xc$, before reaching the destinations, is first passed through a memoryless channels to obtain $\Tp$, resp. $\Tc$. The channel outputs are then given by
  $\Yp = f_1 \left( \Xp,\Tc \right)$ and 
  $\Yc = f_2 \left( \Xc,\Tp \right)$
where $f_u$, $u \in [1:2]$, is a deterministic function which is invertible given $X_u$, or in other words, $\Tp$, resp. $\Tc$, is a deterministic function of $(\Yc,\Xc)$, resp. $(\Yp,\Xp)$.

In the case of source cooperation, also the generalized feedback signal at $\mathsf{Tx2}$ satisfies ${Y}_{\mathsf{F2}} = f_3 \left( \Xc,\Yf \right)$, for some deterministic function $f_3$ that is invertible given $\Xc$ and where $\Yf$ is obtained by passing $\Xp$ through a noisy channel \cite{PVIT11}.

Several outer bounds are known for the IC with bilateral source cooperation~\cite{HostMadsenIT06,PVIT11,TuninettiITA10}, specialized here to the case of unilateral cooperation. 
For an input distribution $P_{X_1,X_2}$:
\begin{subequations}
\label{eq:knownOBgeneralmemo}

\noindent
Case A) For {\em a general memoryless IC with unilateral source cooperation},
the cut-set upper bound \cite{elgamalkimbook} gives
\begin{align}
   \Rp & \leq I \left( \Xp;\Yp,{Y}_{\mathsf{F2}}|\Xc \right) \label{eq:cutset1 a}
\\ \Rp & \leq I \left( \Xp,\Xc; \Yp \right) \label{eq:cutset1 b}
\\ \Rc & \leq I \left( \Xc;\Yc|\Xp \right) \label{eq:cutset2}
\end{align}
and from~\cite{TuninettiITA10} we have
\begin{align}
   \Rp + \Rc & \leq I \left( \Xp;\Yp,{Y}_{\mathsf{F2}}|\Yc,\Xc \right) + I \left( \Xp,\Xc;\Yc \right)
   \label{eq:tuni1}
\\ \Rp + \Rc & \leq I \left( \Xc;\Yc|\Yp,\Xp \right) + I \left( \Xp,\Xc;\Yp \right).
   \label{eq:tuni2}
\end{align}
In~\eqref{eq:cutset1 a}-\eqref{eq:tuni2}, ${Y}_{\mathsf{F2}}$ always appears conditioned on $\Xc$; for the ISD-IC this implies that ${Y}_{\mathsf{F2}}$ can be replaced with $\Yf$.

\noindent
Case B) For {\em a memoryless ISD-IC with unilateral source cooperation that satisfies
$
P_{{Y}_{\mathsf{F2}},\Yp,\Yc|\Xp,\Xc}
=
P_{{\Yp|\Xp,\Xc}}
P_{{Y}_{\mathsf{F2}},\Yc|\Xp,\Xc}
$, i.e., the noises seen by the different source-destination pairs are independent,
}from~\cite{PVIT11} we have
\begin{align}
   \Rp + \Rc
  & \leq 
     H \left( \Yp|\Tp,\Yf \right) - H \left( \Yp|\Tp,\Yf,\Xp,\Xc \right)
\notag\\&+ H \left( \Yc|\Tc,\Yf \right) - H \left( \Yc|\Tc,\Yf,\Xp,\Xc \right) 
\notag\\&+ {I\left(\Yf; \Xp,\Xc| \Tc\right)}.
\label{eq:pv}
\end{align}
The bound in~\eqref{eq:pv} was originally derived by assuming that all noises are independent, which implies $I\left(\Yf; \Xp,\Xc| \Tc\right) \leq I \left( \Yf ; \Xp\right)$. {A straightforward generalization shows that the steps in~\cite{PVIT11} are valid 
even when $P_{{Y}_{\mathsf{F2}},\Yc|\Xp,\Xc}$ is not a product distribution. 
The key step of the proof consists in showing that the following Markov chains hold for all $i\in[1:N]$ for the assumed noise structure:
\begin{align}
&(\Wp, \Tp^{i-1},\Xp^{i}) - (\Tc^{i-1}, \Yf^{i-1}) - (\Tc_{i})         \label{eq:MC with Wp}\\
&(\Wc, \Tc^{i-1},\Xc^{i}) - (\Tp^{i-1}, \Yf^{i-1}) - (\Tp_{i}, \Yf_{i})\label{eq:MC with Wc}.
\end{align}
The proof is not difficult and left out for sake of space.
}

\noindent
Case C) For {\em a memoryless ISD-IC with output feedback ${Y}_{\mathsf{F2}}=\Yc$ and 
with independent noises
$
P_{\Yp,\Yc|\Xp,\Xc}
=
P_{{\Yp|\Xp,\Xc}}
P_{\Yc|\Xp,\Xc},
$
}
from \cite[model (1000)]{SahaiIT2013} we have
\begin{align}
\Rp + 2 \Rc & \leq H \left( \Yc \right) - H \left( \Yc|\Xp,\Xc \right)
\notag\\&+ H \left( \Yc|\Yp,\Xp \right) - H \left( \Yc|\Yp,\Xp,\Xc \right)
\notag\\&          + H \left( \Yp|\Tp \right) - H \left( \Yp|\Tp,\Xp,\Xc \right).
\label{eq:sahai}
\end{align}
\end{subequations}

To the best of our knowledge,~\eqref{eq:sahai} is the only upper bound of the type $\Rp + 2 \Rc$ which is available in the literature, but it is only valid for output feedback.
Our goal in the next section is to derive bounds of the type of~\eqref{eq:sahai} for the class of channels for which~\eqref{eq:pv} is valid.

\section{Novel outer bounds}
\label{sec:outbounds}
In this section we derive two novel outer bounds on the capacity region of the IC with unilateral source cooperation. Our main result is as follows:
\begin{theorem}
\label{th:main th 2r1+r2 and r1+2r2}
For an ISD-IC satisfying
\begin{align*}
&P_{{Y}_{\mathsf{F2}},\Yp,\Yc|\Xp,\Xc}(y_f,y_1,y_2|x_1,x_2)
=\sum_{a,b,c}
P_{\Yf,\Tp|\Xp}(a,b|x_1)
\\&
P_{\Tc|\Xc}(c|x_2)
\delta\big(y_1 \!\!-\!\! f_1(x_1,c)\big)
\delta\big(y_2 \!\!-\!\! f_2(x_2,b)\big)
\delta\big(y_f \!\!-\!\! f_3(x_2,a)\big)
\end{align*}
for some memoryless transition probabilities $P_{\Yf,\Tp|\Xp}, P_{\Tc|\Xc}$ and some injective functions $f_1,f_2,f_3$,
the capacity region is outer bounded by
\begin{align}
2 \Rp + \Rc
&\leq   
      H \left( \Yp \right ) - H \left( \Yp|\Xp,\Xc \right ) \nonumber
\\& + H \left( \Yp|\Tp,\Yf,\Xc \right) - H \left( \Yp|\Tp,\Yf,\Xp,\Xc \right) \nonumber
\\& + H \left( \Yc|\Tc,\Yf \right) - H \left( \Yc|\Tc,\Yf,\Xp,\Xc \right) \nonumber
\\& + I \left( \Yf ;\Xp,\Xc|\Tc \right)
\label{eq:bound2Rp+Rc}
\\
\Rp + 2\Rc
&\leq   
    H \left(\Yc \right)
  - H \left(\Yc | \Xp, \Xc\right)\nonumber
\\&+ 
    H \left( \Yc | \Tc, \Yf, \Xp \right) 
  - H \left( \Yc | \Tc, \Yf, \Xp, \Xc \right)\nonumber
\\&+ 
    H \left( \Yp, \Yf | \Tp\right)
  - H \left( \Yp, \Yf | \Xp, \Xc, \Tp \right)
\label{eq:boundRp+2Rc}
\end{align}
for some input distribution $P_{X_{1},X_{2}}$.
\end{theorem}

\begin{IEEEproof}
By Fano's inequality and by giving side information similarly to \cite{PVIT11}, we have
\begin{align*}
&N(2\Rp+\Rc-3\epsilon_N)
    \leq 
        2I \left( \Wp; \Yp^N \right)
      +  I \left( \Wc; \Yc^N \right) 
\\& \leq 
          I \left( \Wp; \Yp^N \right)
    + I \left( \Wp; \Yp^N, \Tp^N, \Yf^N |\Wc\right)
 \\&   \quad + I \left( \Wc; \Yc^N, \Tc^N, \Yf^N \right) 
\\& \leq 
     H \left( \Yp^N \right )- H \left( \Yp^N, \Tp^N, \Yf^N |\Wp,\Wc \right)
\\& \quad + H \left( \Yp^N, \Tp^N, \Yf^N |\Wc \right) - H \left( \Yc^N, \Tc^N, \Yf^N|\Wc \right)
\\& \quad + H \left( \Yc^N, \Tc^N, \Yf^N \right) - H \left( \Yp^N| \Wp \right ).
\end{align*}
We now analyze each pair of terms.
First pair:
\begin{align*}
  &H \left( \Yp^N \right )- H \left( \Yp^N, \Tp^N, \Yf^N |\Wp,\Wc \right)
\\&\leq \sum_{i\in[1:N]} 
    H \left(\Yp_i \right)
  - H \left(\Yp_{i}, \Tp_{i}, \Yf_{i} | \Xp_{i}, \Xc_{i}\right)
\end{align*}
by using:
the chain rule of the entropy,
the definition of the encoding functions (for the ISD-IC the encoding function $\Xc_{i}(\Wc,{Y}_{\mathsf{F2}}^{i-1})$ is equivalent to $\Xc_{i}(\Wc,\Yf^{i-1})$), the fact that conditioning reduces entropy,
the ISD property of the channel, and
the fact that the channel is memoryless.
Second pair:
\begin{align*}
  &H \left( \Yp^N, \Tp^N, \Yf^N |\Wc \right)
 - H \left( \Yc^N, \Tc^N, \Yf^N|\Wc \right)
\\&= \sum_{i\in[1:N]}
   H \left( \Yp_{i}, \Tp_{i}, \Yf_{i}|\Yp^{i-1}, \Tp^{i-1},  \Yf^{i-1}, \Wc , \Xc^{i}\right)
\\& - \sum_{i\in[1:N]} H \left( \Yc_{i}, \Tc_{i}, \Yf_{i}|\Yc^{i-1}, \Tc^{i-1},  \Yf^{i-1}, \Wc , \Xc^{i}\right)
\\&= \sum_{i\in[1:N]}
   H \left( \Yp_{i}, \Tp_{i}, \Yf_{i}|\Yp^{i-1}, \Tp^{i-1},  \Yf^{i-1}, \Wc , \Xc^{i}\right)
\\& - \sum_{i\in[1:N]}  H \left( \Tp_{i}, \Tc_{i}, \Yf_{i}|\Tp^{i-1}, \Tc^{i-1},  \Yf^{i-1}, \Wc , \Xc^{i}\right)
\\&\leq \sum_{i\in[1:N]}
   H \left( \Tp_{i}, \Yf_{i}|           \Tp^{i-1},  \Yf^{i-1}, \Wc , \Xc^{i}\right)
\\& -  \sum_{i\in[1:N]}  H \left( \Tp_{i}, \Yf_{i}|\Tp^{i-1}, \Tc^{i-1},  \Yf^{i-1}, \Wc , \Xc^{i}\right)
\\&+ \sum_{i\in[1:N]}
   H \left( \Yp_i|\Tp_{i},  \Yf_{i}, \Xc_{i}\right)
\\&-  \sum_{i\in[1:N]}   H \left( \Tc_i|\Tp^{i}, \Tc^{i-1},  \Yf^{i}, \Wc , \Xc^{i}, \Xp^{i}\right)
\end{align*}
where we used:
the chain rule of the entropy,
the definition of the encoding functions,
the ISD property of the channel, 
the fact that conditioning reduces entropy,
the fact that the channel is memoryless, and
since $\Yp$ is a deterministic function of $(\Xp,\Tc)$ that is invertible given $\Xp$;
so finally, by using the Markov chain in \eqref{eq:MC with Wc}, we have
\begin{align*}
  &H \left( \Yp^N, \Tp^N, \Yf^N |\Wc \right)
 - H \left( \Yc^N, \Tc^N, \Yf^N|\Wc \right)
\\&\leq \sum_{i\in[1:N]}
   H \left( \Yp_i|\Tp_{i},  \Yf_{i}, \Xc_{i}\right)
-  H \left( \Yp_i|\Tp_{i},  \Yf_{i}, \Xc_{i}, \Xp_i\right).
\end{align*}
Thrid pair: since
\begin{align*}
& H \left ( \Yp^N|\Wp \right) 
  =   \sum_{i\in[1:N]} H \left ( \Yp_i|\Yp^{i-1},\Wp,{\Xp}^i\right)
\\ &=   \sum_{i\in[1:N]} H \left ( \Tc_i|\Tc^{i-1},\Wp,{\Xp}^i\right)
\\ &\geq \sum_{i\in[1:N]} H \left ( \Tc_i|\Tc^{i-1},\Wp,{\Xp}^i,\Yf^{i-1} \right)
\\ &=   \! \!\!\sum_{i\in[1:N]} 
	\!\!\!H \left ( \Tc_i|\Tc^{i-1},\Yf^{i-1} \right)
\!+\!\underbrace{I \left( \Tc_i;\Wp,\Xp^{i}|\Tc^{i-1},\Yf^{i-1}\right)}_{\text{$=0$ because of~\eqref{eq:MC with Wp}}}
\end{align*}
by using:
the chain rule of the entropy,
the definition of the encoding functions,
the ISD property of the channel, and
the fact that conditioning reduces entropy.
Therefore,
\begin{align*}
  &H \left( \Yc^N, \Tc^N, \Yf^N \right)
  -H \left( \Yp^N|\Wp \right)
\\&\leq \sum_{i\in[1:N]}
  H \left( \Tc_{i}|\Yc^{i-1},\Tc^{i-1},\Yf^{i-1}\right)
 -H \left( \Tc_i|\Tc^{i-1},\Yf^{i-1} \right)
\\& +\sum_{i\in[1:N]} 
  H \left( \Yc_{i},\Yf_{i}|\Yc^{i-1},\Tc^{i},\Yf^{i-1}\right)
\\&\leq \sum_{i\in[1:N]} 0+
  H \left( \Yc_{i},\Yf_{i},|\Tc_i\right).
\end{align*}

By combining everything together, by introducing the time sharing random variable uniformly distributed over $[1:N]$ and independent of everything else, by dividing both sides by $N$ and taking the limit for $N\to\infty$ we get the bound in \eqref{eq:bound2Rp+Rc}. We finally notice that by dropping the time sharing we do not decrease the bound.

By following similar steps as in the derivation of \eqref{eq:bound2Rp+Rc} and by applying the Markov chains in \eqref{eq:MC with Wp} and \eqref{eq:MC with Wc}, it is straightforward to derive the upper bound in~\eqref{eq:boundRp+2Rc} as well.
\end{IEEEproof}


\section{The Gaussian noise channel}
\label{sec:Gaussian}
In this section we specialize the outer bounds to the practically relevant Gaussian noise channel.

\begin{figure}
\centering
\includegraphics[width=0.7\columnwidth]{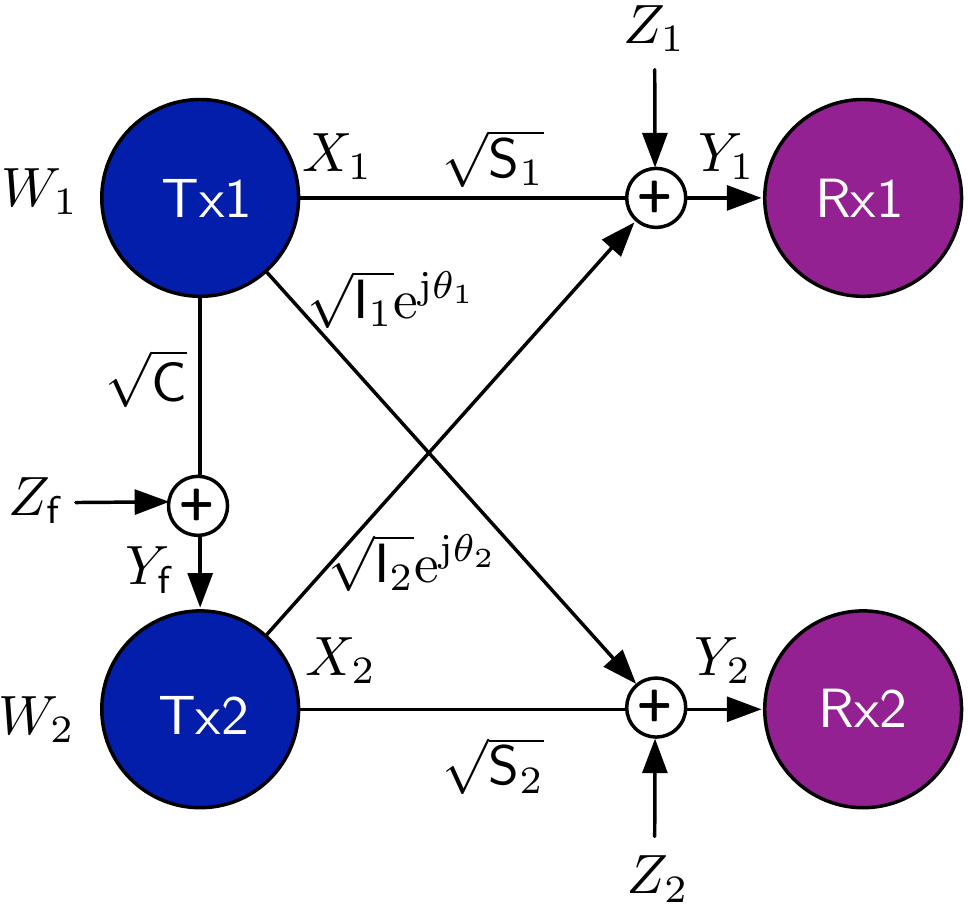}
\caption{The Gaussian IC with unilateral source cooperation.}
\label{fig:channelModel}
\end{figure}

A {\em complex-valued single-antenna full-duplex Gaussian IC with unilateral source cooperation}, shown in Fig.~\ref{fig:channelModel}, is an ISD channel with input/output relationship
\begin{subequations}
\begin{align}
   T_1 &:= \sqrt{\mathsf{I}_1}{\rm e}^{{\rm j}\theta_{1}} \Xp + \Zc, \ \Zc\sim\mathcal{N}(0,1),
\\ T_2 &:= \sqrt{\mathsf{I}_2}{\rm e}^{{\rm j}\theta_{2}} \Xc + \Zp, \ \Zp\sim\mathcal{N}(0,1),
\\ \Yp & = \sqrt{\mathsf{S}_1} \Xp + T_2 : \mathbb{E} \left [ |\Xp|^2 \right ] \leq 1,
\\ \Yc & = T_1 + \sqrt{\mathsf{S}_2} \Xc : \mathbb{E} \left [ |\Xc|^2 \right ] \leq 1,
\\ \Yf & = {Y}_{\mathsf{F2}} = \sqrt{\mathsf{C}} \Xp + \Zf, \ \Zf\sim\mathcal{N}(0,1),
\end{align}
\label{eq:Gmodel}
\end{subequations}
where ${Y}_{\mathsf{F2}}=\Yf$ since $\mathsf{Tx2}$ can remove the contribution of $X_2$ from its received signal. 
The channel gains are constant 
and some are real-valued and non-negative because a node can compensate for the phase of one of its channel gains.
For the assumption under which we derived our outer bounds, we must impose that the noise $\Zp$ is independent of $(\Zc,\Zf)$ (while $(\Zc,\Zf)$ can be arbitrarily correlated).

\subsection{Upper bounds}
The bounds in \eqref{eq:knownOBgeneralmemo}, \eqref{eq:bound2Rp+Rc} and \eqref{eq:boundRp+2Rc} can be evaluated for the Gaussian noise channel in \eqref{eq:Gmodel}. We define  $\mathbb{E} \left [ \Xp \Xc^* \right ] := \rho: |\rho| \in [0,1]$. We also assume that the all noises are independent, which represents a particular case for which our outer bounds hold.
%
By the `Gaussian maximizes entropy' principle, jointly Gaussian inputs exhaust the outer bounds in \eqref{eq:knownOBgeneralmemo}, \eqref{eq:bound2Rp+Rc} and \eqref{eq:boundRp+2Rc}.
We thus start by evaluating each mutual information term in \eqref{eq:knownOBgeneralmemo}, \eqref{eq:bound2Rp+Rc} and \eqref{eq:boundRp+2Rc} by using a jointly Gaussian input.
We then further upper bound (maximize) each mutual information over the input correlation coefficient $\rho:|\rho| \in [0,1]$.
By doing so we obtain: from the cut-set bounds in~\eqref{eq:cutset1 a}-\eqref{eq:cutset2}
\begin{subequations}
\label{eq:knownOBgeneralGaussian}
\begin{align}
   &\Rp  \leq  \log \left( 1 + \Cc + \Sp \right)
\\ &\Rp  \leq  \log\left ( 1+(\sqrt{\Sp} +  \sqrt{\Ic})^2\right )
\\ &\Rc  \leq  \log \left( 1 +  \Sc \right) 
\end{align}
and from the bounds in~\eqref{eq:tuni1}-\eqref{eq:tuni2}
\begin{align}
&\Rp \!+\! \Rc  \leq  \log  \left( 1 \!+\! \frac{ \Sp \!+\!\Cc}{1\!+\! \Ip}  \right) \! + \!  \log  \left( 1\!+\! (\sqrt{\Sc} \!+\! \sqrt{\Ip})^2  \right)
\\ &\Rp \!+\! \Rc  \leq \log  \left( 1 \!+\! \frac{ \Sc}{1\!+\! \Ic}  \right) \!\!+\! \log  \left( 1\!+\! (\sqrt{\Sp} \!+\! \sqrt{\Ic})^2  \right)
\end{align}
and from the bound in~\eqref{eq:pv} (see \cite{PVIT11} under the assumption that all channel gains are larger than one)
\begin{align}
&\Rp + \Rc  \leq\log \left( 1+\Ic + \frac{\Sp}{\Ip} \right)+\log \left( 1+\frac{\Ip}{\Cc} + \frac{\Sc}{\Ic} \right) \nonumber
\\&  \quad  +\log(1+\Cc) +2\log(2).
\end{align}
%

We next evaluate the novel outer bounds in Theorem~\ref{th:main th 2r1+r2 and r1+2r2}, again under the assumption that all channel gains are larger than one, to get
\begin{align}&2\Rp + \Rc \leq  
     \log \left( 1+ \left( \sqrt{\Sp} + \sqrt{\Ic} \right)^2 \right)\nonumber
\\& \quad + \log \left( 1+\Cc\right)  + \log(2) \nonumber
   \\& \quad+ \log \left( 1 + \frac{\Sp}{1+\Ip+\Cc} \right) + \log \left( 1+\frac{\Ip}{\Cc} + \frac{\Sc}{\Ic} \right) \label{eq:2R1 + R2 Gaussian}
\\&\Rp + 2\Rc  \leq \log  \left( 1+ (\sqrt{\Sc} + \sqrt{\Ip})^2 \right)+ \log \left( 1 + \frac{ \Sc}{1+ \Ic}  \right) \nonumber
\\& \quad +\log \left( 1+\Ic + \frac{\Sp+\Cc+\Ic \Cc + 2 \sqrt{\Sp \Ic}}{1+\Ip} \right). \label{eq:R1 + 2R2 Gaussian}
\end{align}
As we shall see in the next section, the new bounds in \eqref{eq:2R1 + R2 Gaussian} and \eqref{eq:R1 + 2R2 Gaussian} are active when the system experiences weak interference and `weak cooperation'.
\end{subequations}


\subsection{Generalized degrees-of-freedom (gDoF) region}
We now focus on the {\em symmetric} Gaussian noise IC with unilateral source cooperation for sake of space.
This channel is parameterized, for some $\snr\geq 1, \alpha\geq 0, \beta\geq 0$, as
\begin{align}
\Sp=\Sc=\snr^1,  \quad
\Ip=\Ic=\inr=\snr^\alpha, \quad
\mathsf{C}=\snr^\beta.
\label{eq:gDoF param}
\end{align}
In particular, we derive the gDoF region, which is an exact capacity characterization in the high-SNR regime $\snr\gg 1$.
The gDoF for the $i$-th user, $i \in [1:2]$, is defined as
\begin{align}
{\gdof}_i
&:= \lim_{\snr\to+\infty} \frac{R_i}{\log(1+\snr)}.
\label{eq:gDoF definition}
\end{align}
This analysis sheds light on an important open question: when unilateral cooperation enables sufficient coordination among the sources for full utilization of the channel resources \cite{suhtse:ICwithfeedback}, i.e., when bounds on $2\Rp+\Rc$ and $\Rp+2\Rc$ are not active.

In \cite{ourICC2014paper}, we showed that, for the Gaussian symmetric IC with unilateral source cooperation, the capacity can be achieved to within a constant gap in strong interference $\alpha \geq 1$, i.e., the interference link is stronger than the direct link, or when $\alpha < 1$ and $\beta \geq \alpha+1$, i.e., the interference is weak and the cooperation link is `sufficiently' strong. This constant gap result implies an exact gDoF region characterization for this set of parameters (the work in \cite{ourICC2014paper} is not restricted to the symmetric case). In these regimes, the capacity region of the Gaussian IC with unilateral source cooperation does not have bounds on $2\Rp+\Rc$ and $\Rp+2\Rc$, similarly to the capacity regions of the classical IC \cite{etw}. In other words, in this regimes the channel resources are fully utilized / there are no `resource holes' in the sense of \cite{suhtse:ICwithfeedback}.

\begin{figure}
\centering
\includegraphics[width=0.8\columnwidth]{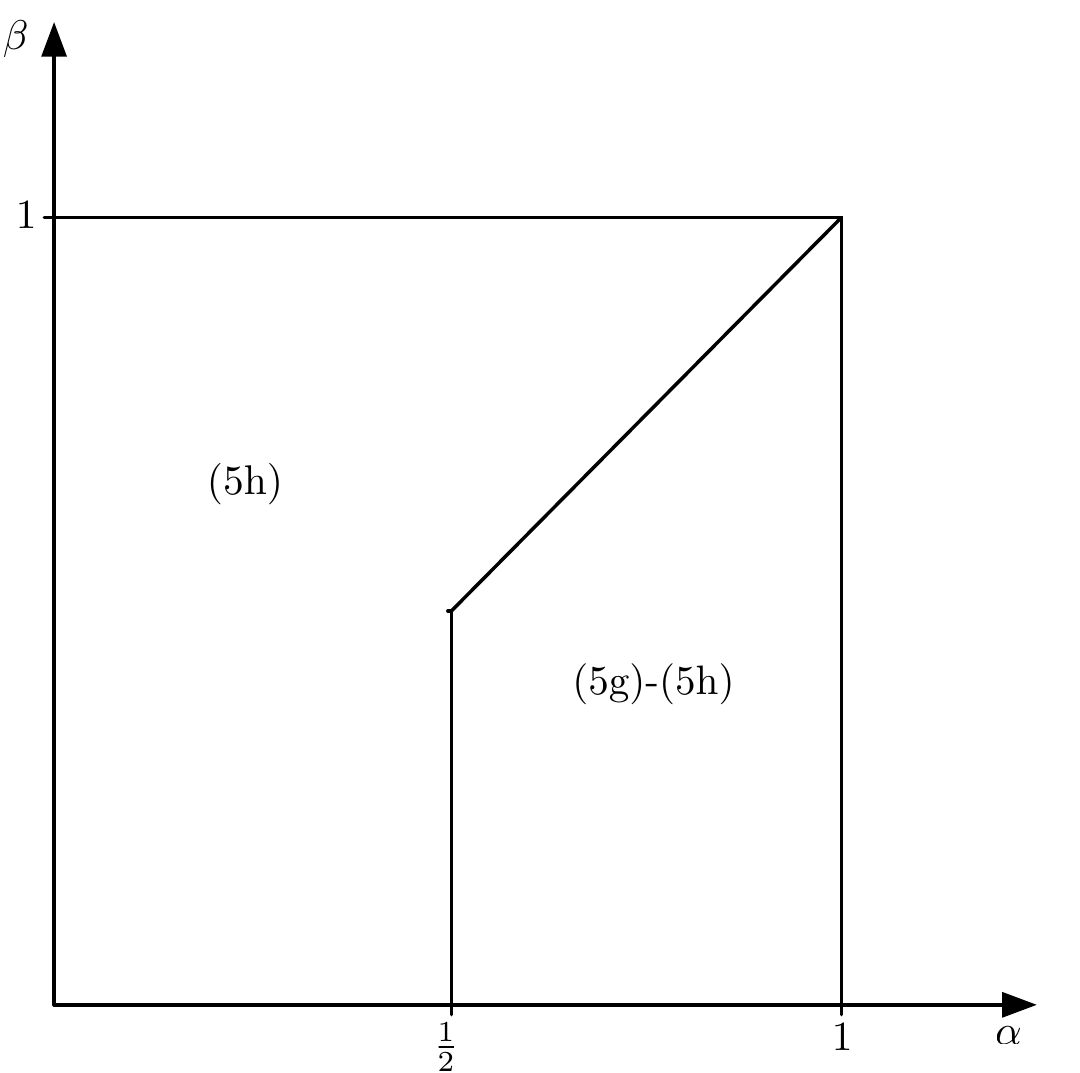}
\caption{Regimes where the bounds in \eqref{eq:2R1 + R2 Gaussian} and \eqref{eq:R1 + 2R2 Gaussian} are active in the gDoF upper bound region  in \eqref{eq:gDoFub}.}
\label{fig:activegDoF}
\end{figure}

Here we focus on a sub-regime left open in \cite{ourICC2014paper}, namely $\alpha < 1$ and $\beta \leq 1$, i.e., when the direct links are stronger than the interfering links and the cooperation link.
By using the definition in \eqref{eq:gDoF definition}, from the upper bound region in \eqref{eq:knownOBgeneralGaussian}, we obtain that the gDoF region of the Gaussian IC with unilateral cooperation, when $\alpha < 1$ and $\beta \leq 1$, is upper bounded by
\begin{subequations}
\begin{align}
   {\gdof}_1 & \leq 1 \label{eq:cutR1}
\\ {\gdof}_2 & \leq 1 \label{eq:cutR2} 
\\ {\gdof}_1 \!+\! {\gdof}_2 & \leq  2-\alpha \label{eq:Tuninetti}
\\ {\gdof}_1 \!+\! {\gdof}_2 & \leq \max \left \{ \alpha,1-\alpha \right \} + \max \left \{ \alpha, 1+\beta -\alpha\right \}  \label{eq:PV}
\\ 2{\gdof}_1 \!+\! {\gdof}_2 & \leq \! 1 \!+\!  [1 \!-\! \max \left \{\alpha,\beta \right \}]^+  \!+\! \max \left \{ \alpha, 1\!-\!\alpha\!+\!\beta\right \} \label{eq:2d1+d2}
\\ {\gdof}_1 \!+\! 2{\gdof}_2 & \leq 2-\alpha +\max \left \{ \alpha,\beta,1-\alpha \right \}. \label{eq:d1+2d2}
\end{align}
\label{eq:gDoFub}
\end{subequations}
From the gDoF region above interesting insights can be drawn on when the capacity of the IC with unilateral source cooperation has bounds on $2\Rp+\Rc \ / \ \Rp+2\Rc$. Fig. \ref{fig:activegDoF} shows which one among the two new bounds in \eqref{eq:2R1 + R2 Gaussian}-\eqref{eq:R1 + 2R2 Gaussian}  are active.  From Fig. \ref{fig:activegDoF} we observe that both bounds are active whenever $\alpha \geq \max \left\{\frac{1}{2},\beta \right\}$, while in the other case only $\Rp+2\Rc$ is active. Moreover, we notice that, in weak interference, i.e., $\alpha<1$, and with $\beta\leq [2\alpha-1]^+$, the gDoF region in \eqref{eq:gDoFub} is the same as that of the classical non-cooperative IC \cite{etw}. Therefore, for this set of parameters the gDoF outer bound region in \eqref{eq:gDoFub} is achievable to within a constant gap (derivation straightforward and not shown here for sake of space). For the other parameter regimes, developing strategies that achieve the gDoF upper bound in \eqref{eq:gDoFub} is an important open problem, which is object of current investigation.


\section{Conclusions}
\label{sec:conclusion}
In this work we studied the two-user IC with unilateral source cooperation where one source overhears the other source through a noisy in-band link. Our major
contribution was to develop two novel outer bounds of the type $2\Rp+\Rc/\Rp+2\Rc$ on the capacity of this system. These bounds were first derived for the injective semi-deterministic channel and then specialized to the Gaussian case. The symmetric, i.e., equally strong interfering links and direct link, Gaussian channel was investigated in the high SNR regime, in order to highlight under which channel conditions unilateral cooperation enables sufficient coordination among the sources such that the $2\Rp+\Rc/\Rp+2\Rc$-type bounds are not active.

%

\bibliographystyle{IEEEtran}
\bibliography{ISITBib}

\begin{thebibliography}{10}
\providecommand{\url}[1]{#1}
\csname url@samestyle\endcsname
\providecommand{\newblock}{\relax}
\providecommand{\bibinfo}[2]{#2}
\providecommand{\BIBentrySTDinterwordspacing}{\spaceskip=0pt\relax}
\providecommand{\BIBentryALTinterwordstretchfactor}{4}
\providecommand{\BIBentryALTinterwordspacing}{\spaceskip=\fontdimen2\font plus
\BIBentryALTinterwordstretchfactor\fontdimen3\font minus
  \fontdimen4\font\relax}
\providecommand{\BIBforeignlanguage}[2]{{%
\expandafter\ifx\csname l@#1\endcsname\relax
\typeout{** WARNING: IEEEtran.bst: No hyphenation pattern has been}%
\typeout{** loaded for the language `#1'. Using the pattern for}%
\typeout{** the default language instead.}%
\else
\language=\csname l@#1\endcsname
\fi
#2}}
\providecommand{\BIBdecl}{\relax}
\BIBdecl

\bibitem{HostMadsenIT06}
A.~Host-Madsen, ``Capacity bounds for cooperative diversity,'' \emph{IEEE
  Trans. on Info. Theory}, vol.~52, no.~4, pp. 1522 --1544, April 2006.

\bibitem{PVIT11}
V.~Prabhakaran and P.~Viswanath, ``Interference channels with source
  cooperation,'' \emph{IEEE Trans. on Info. Theory}, vol.~57, no.~1, pp. 156
  --186, Jan. 2011.

\bibitem{TuninettiITA10}
D.~Tuninetti, ``An outer bound region for interference channels with
  generalized feedback,'' in \emph{Information Theory and Applications Workshop
  (ITA), 2010}, Feb. 2010, pp. 1--5.

\bibitem{YANG-TUNINETTI}
S.~Yang and D.~Tuninetti, ``Interference channel with generalized feedback
  (a.k.a. with source cooperation): Part i: Achievable region,'' \emph{IEEE
  Trans. on Info. Theory}, vol.~57, no.~5, pp. 2686 --2710, May 2011.

\bibitem{YangHighCoop}
------, ``Interference channels with source cooperation in the strong
  cooperation regime: Symmetric capacity to within 2 bits/s/hz with dirty paper
  coding,'' in \emph{Asilomar 2011}, Nov. 2011, pp. 2140 --2144.

\bibitem{MirmohseniIT2012}
M.~Mirmohseni, B.~Akhbari, and M.~Aref, ``On the capacity of interference
  channel with causal and noncausal generalized feedback at the cognitive
  transmitter,'' \emph{IEEE Trans. on Info. Theory}, vol.~58, no.~5, pp.
  2813--2837, May 2012.

\bibitem{ourICC2014paper}
M.~Cardone, D.~Tuninetti, R.~Knopp, and U.~Salim, ``On the capacity of
  full-duplex causal cognitive interference channels to within a constant
  gap,'' \emph{to appear in IEEE International Conference on Communications
  (ICC), 2014 extended version available on arXiv:1207.5319v3}, 2014.

\bibitem{suhtse:ICwithfeedback}
C.~Suh and D.~Tse, ``Feedback capacity of the gaussian interference channel to
  within 2 bits,'' \emph{IEEE Trans. on Info. Theory}, vol.~57, no.~5, pp.
  2667--2685, 2011.

\bibitem{SahaiIT2013}
A.~Sahai, V.~Aggarwal, M.~Yuksel, and A.~Sabharwal, ``Capacity of all nine
  models of channel output feedback for the two-user interference channel,''
  \emph{IEEE Trans. on Info. Theory}, vol.~59, no.~11, pp. 6957--6979, Nov.
  2013.

\bibitem{TandonPoor}
S.-Q. Le, R.~Tandon, M.~Motani, and H.~Poor, ``The capacity region of the
  symmetric linear deterministic interference channel with partial feedback,''
  in \emph{in 50th Annual Allerton Conference on Communication, Control, and
  Computing}, 2013, pp. 1864--1871.

\bibitem{TelatarTseISIT2007}
I.~Telatar and D.~Tse, ``Bounds on the capacity region of a class of
  interference channels,'' in \emph{IEEE International Symposium on Information
  Theory (ISIT), 2007}, 2007, pp. 2871--2874.

\bibitem{elgamalkimbook}
A.~E. Gamal and Y.-H. Kim, \emph{Network Information Theory}.\hskip 1em plus
  0.5em minus 0.4em\relax Cambridge U.K.: Cambridge Univ. Press,, 2011.

\bibitem{etw}
R.~Etkin, D.~Tse, and H.~Wang, ``Gaussian interference channel capacity to
  within one bit,'' \emph{IEEE Trans. on Info. Theory}, vol.~54, no.~12, pp.
  5534 --5562, Dec. 2008.

\end{thebibliography}
\end{document}